\documentclass{iau} 

\usepackage{graphicx}
\usepackage{xcolor}
\usepackage[hyphens]{url}
\usepackage{hyperref}
\hypersetup{
     colorlinks  = true,
     linkcolor   = blue, 
     anchorcolor = BrickRed,
     citecolor   = blue,
     filecolor   = blue,
     menucolor   = blue,
     runcolor    = cyan,
     urlcolor    = blue
}
\usepackage[authoryear]{natbib}
\usepackage{comment}

\pubyear{2019}
\volume{355} 
\jname{The Realm of the Low-Surface-Brightness Universe}
\editors{D. Valls-Gabaud, I. Trujillo \& S. Okamoto, eds.}

\title[The LSB Universe: a new frontier in galaxy evolution studies] 
{The low-surface-brightness Universe: a new frontier in the study of galaxy evolution}

\author[Sugata Kaviraj]   
{Sugata Kaviraj$^1$}

\affiliation{$^1$ Centre for Astrophysics Research, University of Hertfordshire, College Lane, Hatfield AL10 9AB, UK \\email: {\tt s.kaviraj@herts.ac.uk}}

\begin{document}

\maketitle

\begin{abstract}
New and forthcoming deep-wide surveys, from instruments like the Hyper-Suprime-Cam, LSST and EUCLID, are poised to revolutionize our understanding of galaxy evolution, by revealing aspects of galaxies and their environments that are largely invisible in past wide-area datasets. These surveys will open up the realm of low-surface-brightness (LSB) and dwarf galaxies -- which dominate the galaxy number density -- for the first time at cosmological distances. They will also reveal key, unexplored LSB structures which put stringent constraints on our structure-formation paradigm, such as merger-induced tidal features (which encode galaxy assembly histories) and intra-cluster light (which can dominate the baryonic mass budget of galaxy clusters). However, exploitation of these revolutionary new datasets will require us to address several data-analysis challenges. Data-processing pipelines will have to preserve LSB structures, which are notoriously susceptible to sky over-subtraction and shredding by de-blenders. Analysis of the prodigious data volumes will need to incorporate machine-learning (in particular unsupervised techniques), to augment or even replace traditional methods. Cosmological simulations, which are essential for a statistical understanding of the physics of galaxy evolution, will require mass and spatial resolutions that are high enough to resolve LSB/dwarf galaxies and LSB structures. And finally, the estimation of physical properties (e.g. stellar masses and star formation rates) will require reliable redshift information. Since it is unlikely that even the next generation of spectrographs will provide complete spectral coverage in the LSB/dwarf regime outside the nearby Universe, photometric redshifts may drive much of the science from these datasets. It is necessary, therefore, that the accuracy of these redshifts is good enough (e.g. $<$10 per cent) to enable statistical studies of galaxy evolution in the LSB/dwarf regime. Here, I outline the tremendous discovery potential of new and forthcoming deep-wide surveys and give examples of techniques which will enable us to solve the data-analysis challenges outlined above.

\end{abstract}

\firstsection 

\section{The low-surface-brightness Universe: a key, unexplored domain}
Our statistical understanding of how the Universe evolves is strongly determined by the objects and structures that are brighter than the surface-brightness limits of wide-area surveys. The past few decades have produced a revolution in both the way astronomical data are gathered and the way they are interpreted using theoretical models. Wide-area surveys, like the SDSS \citep{Abazajian2009}, have enabled us to study the statistical properties of the galaxy population, based on hundreds of thousands of objects. And the convergence of such datasets with simulations in cosmological volumes \citep[e.g.][]{Somerville2012,Dubois2014,Vogelsberger2014,Schaye2015,Kaviraj2017} has allowed us to elucidate the principal processes that drive galaxy evolution over cosmic time. 

However, while huge strides have been made in comprehending galaxy evolution, our understanding is naturally constrained by aspects of the Universe that are actually observable in past surveys. The completeness of galaxies in surveys like the SDSS decreases rapidly for surface brightnesses fainter than $\sim$24.5 mag arcsec$^{-2}$ \citep[e.g.][]{Kniazev2004,Driver2005}. This renders much of the low-surface-brightness (LSB) regime \citep[e.g.][]{vd2014,Greco2018}, which includes the dwarf galaxy population, inaccessible at cosmological distances \citep[e.g.][]{Martin2019}. However, both theory \citep[e.g.][]{Martin2019} and observation \citep[e.g.][]{Dalcanton1997} indicate that the bulk of the galaxy population actually resides in the LSB/dwarf regime. For example, $\sim$50 ($\sim$85) per cent of galaxies down to $10^8$ ($10^7$) M$_{\odot}$ inhabit this regime \citep[see Table 2 in][]{Martin2019} . This has two important consequences. First, our empirical knowledge of galaxy evolution is based on potentially the `tip of the iceberg' i.e. a small subset of high surface-brightness (HSB) systems. Second, and more importantly, our understanding of the physics of galaxy evolution is predicated on a small subset of the galaxy population. Both of these facts render our understanding of how the Universe evolves (potentially highly) incomplete. Not surprisingly, many of the well-known tensions between theory and observation are in the LSB/dwarf regime, e.g. the apparent underproduction of dwarfs seen in simulations \citep{Moore1998} and the core-cusp problem \citep[e.g.][]{Navarro1996}. 

Apart from incompleteness in the galaxy population itself, there are key LSB components of even HSB galaxies that are largely invisible in past wide-area surveys (at least outside the local Universe), but which offer strong constraints on our theoretical paradigm. Two examples are merger-induced LSB tidal features and intra-cluster light (ICL). Tidal features encode the assembly histories of galaxies and constrain our structure-formation model. However, the surface-brightness of tidal features is a strong function of merger mass ratio. Given that low-mass galaxies far outnumber their massive counterparts, most mergers involve low mass ratios (i.e. are `minor' mergers), which produce faint tidal features that are largely undetectable in past wide-area surveys \citep{Kaviraj2010,Duc2015}. Nevertheless, both theory \cite[e.g.][]{Oser2012,Martin2018} and observation \citep[e.g.][]{Kaviraj2014} suggest that minor mergers are key drivers of galaxy evolution, making the analysis of LSB tidal features an essential component of our galaxy-evolution effort. In a similar vein, ICL is a significant component of galaxy clusters, which are important tests of our cosmological model. Since the ICL contributes a significant (and sometimes a dominant fraction) of the baryonic mass budget of clusters \citep[e.g.][]{Burke2012}, the utility of clusters is closely linked to our ability to detect and characterize the ICL over cosmic time. 


\section{Exploitation of new deep-wide surveys: challenges and solutions}

The advent of a new generation of deep-wide surveys is set to transform our view of the optical Universe. These surveys will not only enable, for the first time, statistical studies at unprecedented surface-brightness limits, but also provide optical counterparts and ancillary information (e.g. photometric redshifts, stellar masses, star formation rates etc.) for nearly all future surveys at other wavelengths. 

For example, the Hyper Suprime-Cam Subaru Strategic Program (HSC-SSP), an ongoing survey which is already providing data, will offer, by 2021, $\sim$1400 deg$^2$ in $grizy$ down to a depth of r$\sim$26 ($\sim$4 mags deeper than the SDSS), with a median seeing of 0.6". LSST \citep{Robertson2019}, a ground-breaking instrument that was the top ranked ground-based facility in the 2010 NSF Decadal survey, will provide, in its commissioning phase, $\sim$2000 deg$^2$ in $(u)grizy$ to a depth of r$\sim$26.5 and around 200 deg$^2$ to r$\sim$28, both with a median seeing of 0.7". LSST commissioning data is expected by 2022. The full LSST Wide survey will offer $\sim$18,000 deg$^2$ to a depth of r$\sim$28 in the early 2030s. Finally, EUCLID (to be launched in 2022) will provide, via its Wide survey, $\sim$15,000 deg$^2$ down to r$\sim$24.5, at a resolution similar to that of the Hubble Space Telescope. Together, these surveys will provide a database of billions of galaxies and LSB structures, that will facilitate an unprecedented empirical exploration of the LSB Universe and enable us to use the astrophysical results from this largely unexplored regime to constrain our theoretical paradigm. 

\begin{figure}
\centering
\includegraphics[width=0.8\textwidth]{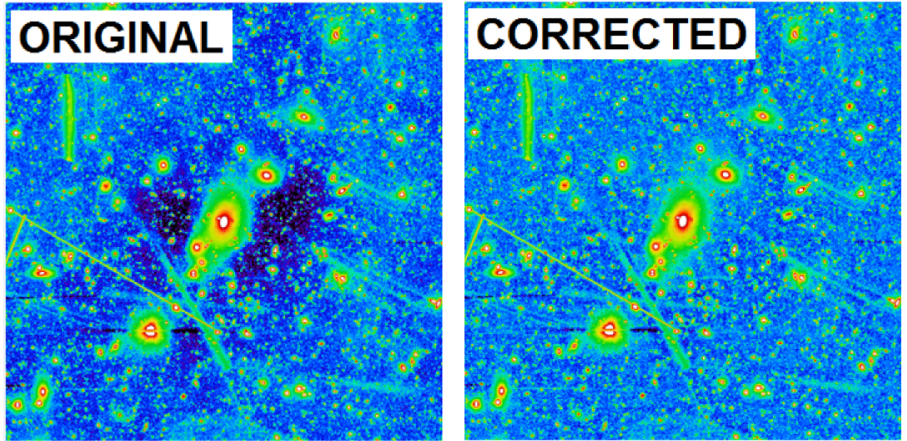}
\caption{Left: HSC-SSP DR1 i-band image of the XCS 35 cluster. Sky over-subtraction is visible around bright and extended sources. Right: A non-parametric mixed model flux threshold correction (Kelvin et al. in preparation), which successfully restores over-subtracted regions. See text in Section 2.1 for details.}
\label{fig:LSB preservation}
\end{figure}

However, notwithstanding their transformational nature, there are several data-analysis challenges that must be addressed before these datasets can be fully utilized. Four key challenges are as follows (this list is not exhaustive, but offers examples of broad areas in which the community will need to invest, in preparation for these surveys):

\begin{itemize}

    \item LSB structures are acutely sensitive to sky over-subtraction and shredding by de-blenders. Preservation of LSB flux in survey images is, therefore, a key requirement of the data-processing pipelines employed by deep-wide surveys. 

    \item The sheer volume of data expected from such surveys will demand novel analysis techniques. For example, the raw data output of LSST is 20 TB per night (over roughly a ten year period). Automation, delivered via artificial intelligence and machine-learning methods, will be required to augment, and perhaps even replace, traditional analysis techniques.
    
    \item The interpretation of these wide-area surveys will ideally require hydro-dynamical simulations in cosmological volumes. However, the current generation of simulations, like Horizon-AGN \citep[e.g.][]{Dubois2014}, Illustris \citep[e.g.][]{Vogelsberger2014} and EAGLE \citep[e.g.][]{Schaye2015}, lack the mass and spatial resolution to fully resolve LSB/dwarf galaxies (particularly those with stellar masses lower than $\sim$10$^9$ M$_{\odot}$) and LSB structures around low-mass galaxies. New simulations with higher mass/spatial resolution will be necessary for detailed studies of galaxies in the LSB/dwarf regime (although previous generations of simulations and semi-analytical models will remain important, since the increase in mass/spatial resolution comes at the expense of simulation box sizes). 

    \item The derivation of physical quantities, e.g. stellar masses, rest-frame colours and star formation rates, requires reliable redshift information. While planned spectroscopic campaigns \citep[e.g. WAVES,][]{Driver2016} will cover some of the footprint of surveys like those from the LSST, it is unlikely that full spectroscopic coverage will be available in the LSB/dwarf regime outside the nearby Universe. Photometric redshifts will, therefore, underpin significant amounts of science from these surveys and will require accuracies that are high enough to enable statistical galaxy-evolution studies using these datasets.

\end{itemize}

The following sections explore these data-analysis challenges and potential solutions further. 


\subsection{Preserving LSB structures in deep-wide survey imaging}

Given their faintness, LSB structures are sensitive to the modelling and subtraction of the sky. In particular, they can be affected by sky over-subtraction in pipelines that are optimized for the detection of relatively HSB structures. Since past wide-area surveys have primarily sampled the HSB Universe, it is not uncommon for existing pipelines to over-subtract the sky, removing a significant fraction of the LSB flux. 
Figure \ref{fig:LSB preservation} shows an example of this issue. The left-hand panel shows the cluster XCS 35 in an HSC-SSP DR1 \citep{Tanaka2018} $i$-band image. The pipeline over-subtracts the sky around bright and extended sources, which is likely to make ICL studies difficult, since some of the ICL flux has clearly been removed. The right-hand panel describes a possible solution, that uses a non-parametric approach to restore over-subtracted regions. In this case, 2D Sersic models are fitted to sources to characterize the expected flux in the wings of each object. The residual between the model and science maps can then be used to define a flux threshold below which over-subtraction becomes a problem and values below this threshold can be added back to restore the over-subtracted regions (Kelvin et al., in preparation). 

While such bespoke treatments can be used to restore regions of LSB flux in individual images, the challenge is to implement such techniques (or adaptations of existing algorithms, e.g. \citet{Akhlaghi2015}) to augment or alter the \emph{standard} pipelines that bulk-process data from surveys like LSST (see the IAC Strip 82 lecgacy project for an example of such an exercise on SDSS Stripe82 data, \citet{Trujillo2017}).


\subsection{Machine-learning methods for deep-wide surveys: an example based on the morphological analysis of galaxies}

The prodigious data volumes expected from new and future surveys may render traditional methods inadequate, making it important to augment or even replace classical methods using artificial intelligence and machine-learning techniques. 

\begin{figure}
\centering
\includegraphics[width=0.8\textwidth]{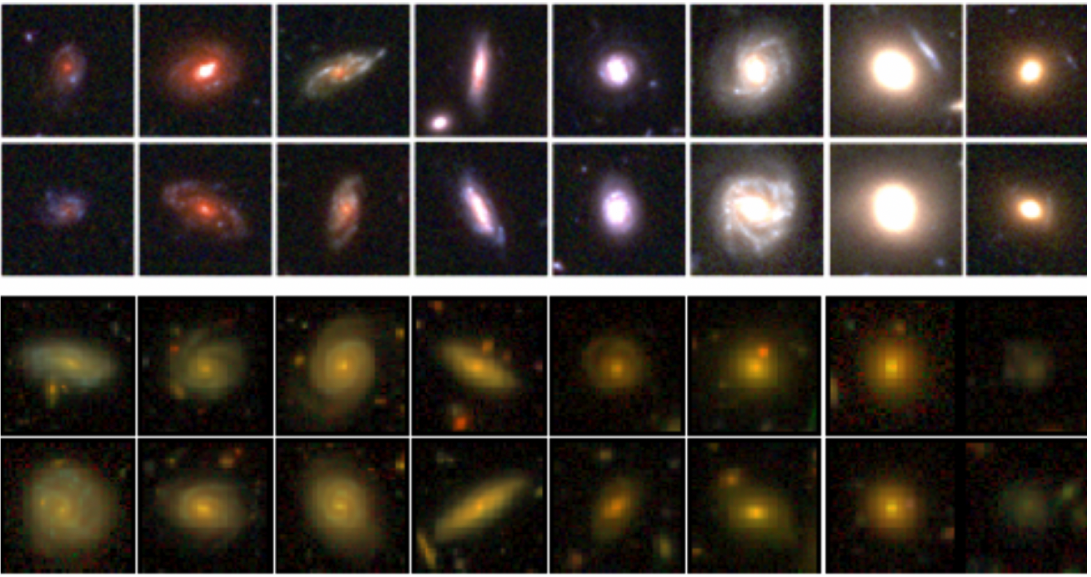}
\caption{Top two rows: The unsupervised machine-learning algorithm described in Section 2.2 implemented on HST data produces clean separation of objects that are composed of pixels with different properties e.g. colour and texture. Bottom two rows: An implementation on ground-based images, from the Ultra-deep layer of the HSC-SSP DR1 \citep{Martin2020}. The HSC-SSP DR1 Ultradeep images have similar resolution to LSST and a depth that is similar to one of the LSST commissioning surveys (see Section 1) and its final Wide survey. This figure may look better on screen than in print.} 
\label{fig:ML}
\end{figure}

Consider, as an example, the morphological analysis of galaxies. Galaxy morphology is a fundamental parameter, that is critical not only for the full spectrum of galaxy evolution studies, but also for a plethora of science across observational cosmology, e.g. as a prior for photometric redshift pipelines and as contextual data for transient lightcurve classifications. Instruments like the HSC, LSST and EUCLID offer an unprecedented opportunity for the morphological analysis of galaxies (and structures like tidal features), down to lower stellar masses and fainter surface brightnesses than ever before. Indeed, the morphological mix of galaxies in the LSB regime is likely to be different from the Hubble sequence that describes the HSB population and remains unquantified. However, the unprecedented data volumes from deep-wide surveys may make traditional methods of classification, like visual inspection \citep{Willett2015}, intractable, even using massively-distributed systems like Galaxy Zoo \citep{Lintott2011}. The short cadence of surveys like LSST may pose an additional hurdle, because repeatedly producing training sets that are required for supervised machine-learning also becomes impractical. 

\begin{figure}
\begin{tabular}{cc}
\includegraphics[width=0.8\textwidth]{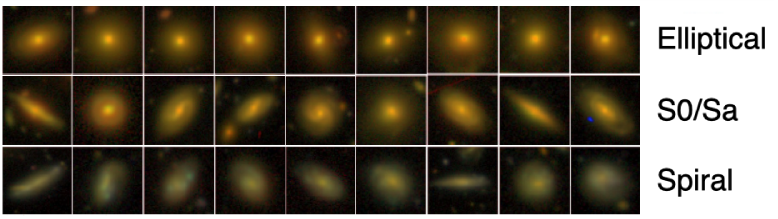}\\
\includegraphics[width=0.41\textwidth]{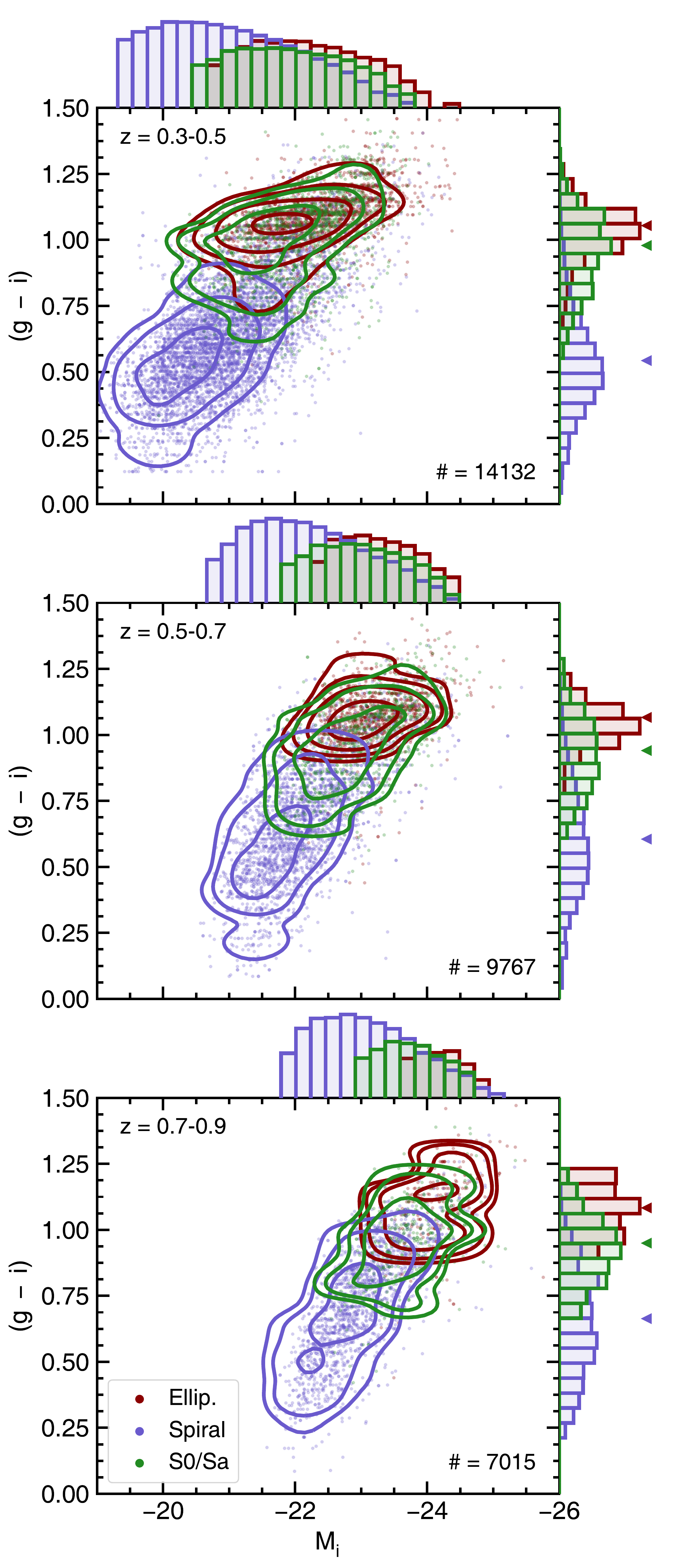}
\includegraphics[width=0.41\textwidth]{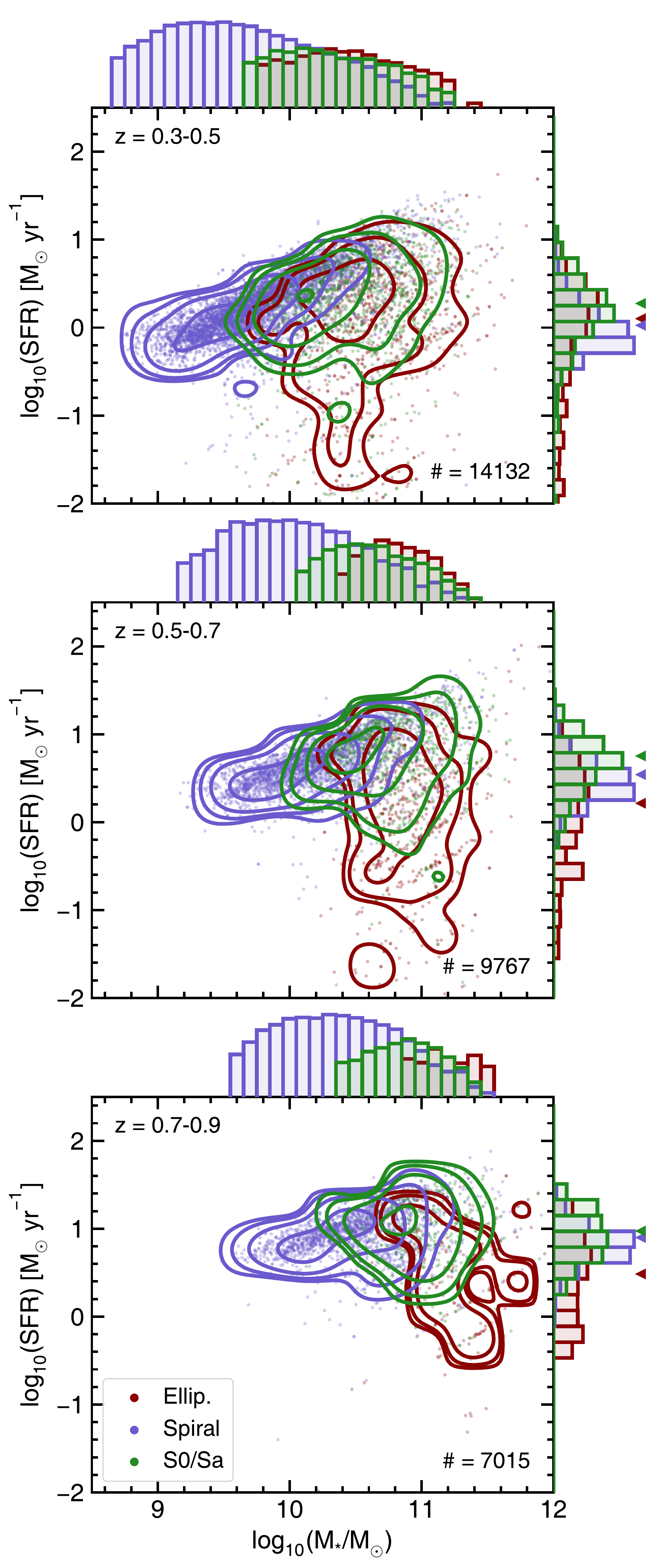}
\end{tabular}
\caption{Physical properties of galaxies -- rest-frame colours and absolute magnitudes (left) and the star-formation main sequence (right) -- in broad morphological classes (ellipticals, S0/Sa and discs; shown in the top panel) in the low and intermediate redshift Universe ($z<1$). Galaxies classified into these broad morphological classes reproduce known trends in galaxy properties as a function of morphology.}
\label{fig:ML2}
\end{figure}

\begin{figure}
\centering
\includegraphics[width=\textwidth]{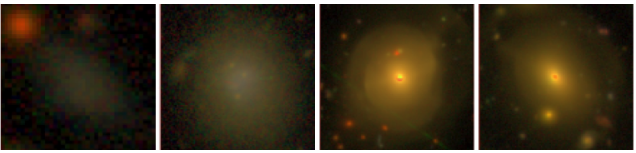}
\caption{Examples of two morphological clusters from the unsupervised machine-learning algorithm described in Section 2.2 which involve LSB galaxies or structures. The first two images (from the left) show examples of objects from a morphological cluster populated by relatively massive LSB dwarfs. The remaining images show examples of objects from a morphological cluster populated by elliptical galaxies that exhibit LSB shells (which are indicative of recent minor mergers). This figure may look better on screen than in print.}
\label{fig:ML3}
\end{figure}

\begin{figure}
\centering
\includegraphics[width=\textwidth]{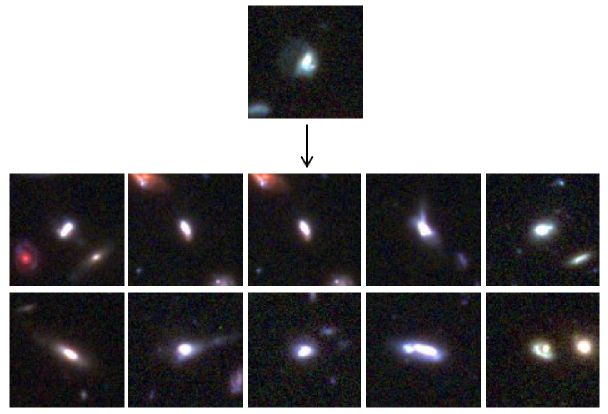}
\caption{An example of a `similarity search' where an archetypal object can be used to find similar systems in survey images. The similarity is established via the galaxy feature vectors. In this case an object with known tidal features has been used to find similar tidally-disturbed objects in HST CANDELS \citep{Grogin2011} images (see text in Section 2.2 for details of how the feature vectors are defined). This figure may look better on screen than in print.}
\label{fig:ML4}
\end{figure}

For such surveys, \textit{unsupervised} machine learning (UML) offers an attractive route to morphological analysis. An effective UML algorithm can autonomously group similar objects (e.g. galaxies), in survey images, thus compressing an arbitrarily large galaxy population into a small number of `morphological clusters', which can then be benchmarked via visual classification. Figure \ref{fig:ML} shows an example of such a UML algorithm, which has been successfully implemented on both space-based HST \citep{Hocking2018} and ground-based HSC-SSP \citep{Martin2020} images. The technique extracts image patches from multi-band data, each of which is transformed into a rotationally-invariant representation of a small region of the survey data, efficiently encoding colour, intensity and spatial frequency information. Utilizing growing neural gas and hierarchical clustering algorithms, it then groups the patches into a library of patch types, based on their similarity, and assembles `feature vectors' for each object, which describe the frequency of each patch type. A $k$-means algorithm is then used to separate objects into 160 morphological clusters, based on the similarity of their feature vectors \citep{Martin2020}.

These clusters can then be benchmarked via visual inspection, both to establish the morphology of the average object and the morphological purity of the cluster. If the purity of the clusters is high, then the benchmarking exercise reduces to visual inspection of 160 \textit{clusters}, rather than hundreds of thousands of individual galaxies, making the problem tractable even for individual researchers. Figure \ref{fig:ML} shows a few morphological clusters produced by the algorithm (individual columns represent morphological clusters). The top two rows show an implementation on HST data, while the bottom two rows show an implementation on ground-based HSC data. 

The discrimination between broad morphological classes (e.g. elliptical galaxies, S0/Sa systems and spiral galaxies) is very accurate, with high purity within morphological clusters. Galaxies classified into these morphological classes reproduce known trends in physical properties (e.g. stellar masses, absolute magnitudes, rest-frame colours and star formation rates) with redshift at $z<1$ (Figure \ref{fig:ML2}). Note that it is difficult to go beyond this epoch, given the ground-based resolution of the HSC images. While the algorithm has not been optimized to find LSB galaxies and structures, it shows promise in identifying galaxies with faint LSB shells and relatively massive dwarfs in the HSC-SSP DR1 images (Figure \ref{fig:ML3}). Finally, since the characteristics of different objects are encoded by their feature vectors, comparison of these vectors makes it possible to find similar systems given an archetype. Figure \ref{fig:ML4} shows an example, where an object with known LSB tidal features can be used to identify other similar objects in the galaxy sample. Readers are directed to \citet{Martin2020} for more details of this algorithm.

As these examples show, the potential of UML is significant in the context of future surveys. The algorithms outlined here (and others like them) will be able to tackle data from both space- (e.g. EUCLID) and ground-based (e.g. LSST) instruments. While they will need to be optimized for the characteristics of specific surveys, in particular their surface-brightness limits, UML algorithms will provide an essential platform for performing automated data analysis of new and forthcoming datasets. 


\subsection{High-resolution hydro-dynamical simulations in cosmological volumes}

\begin{figure}
\centering
\includegraphics[width=0.8\textwidth]{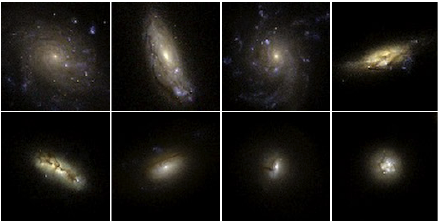}
\caption{Examples of galaxies in the New Horizon simulation (Dubois et al. in preparation) which offers stellar mass and maximum spatial resolutions of $10^4$~M$_{\odot}$ and 40 pc respectively, in a volume with a radius of 10 Mpc. The galaxies shown are in the mass range $10^7$~M$_{\odot}$ $<$ M$_*$ $<$ $10^{10}$~M$_{\odot}$. We show mock $gri$ colour images at LSST resolution for nearby galaxies. This figure may look better on screen than in print.}
\label{fig:NH}
\end{figure}

The recent advent of hydro-dynamical simulations in cosmological volumes (e.g. Horizon-AGN, EAGLE and Illustris), in which baryons and dark matter are evolved self-consistently, has opened up a powerful new set of tools with which to understand the physics of galaxy evolution. 
Nevertheless, both the mass and spatial resolutions of such simulations are not ideally suited to the LSB/dwarf regime. For example, the stellar mass and spatial resolutions of Horizon-AGN are $4\times10^6$~M$_{\odot}$ and 1 kpc respectively. Since 100 star particles are required for detecting structures, the effective \textit{galaxy} mass resolution is $\sim$10$^{8.5}$ M$_{\odot}$, making the dwarf regime largely inaccessible. Furthermore, a 1 kpc resolution does not fully resolve the scale heights of disks, especially in low-mass galaxies (the Milky Way scale height, for comparison, is $\sim$300 pc). Finally, while LSB tidal features do appear around massive galaxies in these simulations, the mass resolution does not lend itself well to the detection of such features around lower mass systems. 

A new generation of higher-resolution cosmological hydro-dynamical simulations offers an opportunity both to make statistical predictions in the LSB/dwarf regime and to better resolve the internal properties of galaxies. An example is the New Horizon simulation (Dubois et al. in preparation), which offers a mass and a maximum spatial resolution of $10^4$~M$_{\odot}$ and 40 pc respectively, in a volume with radius 10 Mpc (Figure \ref{fig:NH}). New Horizon, together with other high-resolution cosmological simulations like Illustris-TNG50 \citep{Pillepich2019} and cosmological `zoom-ins' \citep[e.g.][]{Hopkins2014,Wang2015}, will provide valuable theoretical counterparts to the datasets expected from instruments like LSST and EUCLID. It is worth noting, however, that the increase in resolution typically comes at the expense of the box size of the simulation. This restricts the range of environments probed by the simulations and means that they do not contain rare objects, like the most massive galaxies. As a result, these high-resolution simulations will likely be used in conjunction with previous generations of hydro-dynamical simulations and also semi-analytical models, which lend themselves well to the rapid, phenomenological exploration of parameter space. 


\subsection{Statistical studies using photometric redshifts in the deep-wide survey era}

While new and forthcoming surveys bring the prospect of probing fainter (and therefore lower-mass) galaxies, our ability to study galaxy evolution depends on the derivation of physical properties, such as stellar mass, star formation rate and rest-frame colours. This, in turn, requires reliable redshift information. While planned spectroscopic surveys will cover limited fractions of the footprints of forthcoming surveys like those from the LSST, spectroscopic coverage of the entire footprints down into the LSB/dwarf regime is unlikely, at least in the near future. It is, therefore, interesting to consider whether photometric redshifts alone could be used to perform statistical studies of galaxies using such datasets. The depth of the data from deep-wide surveys should afford better photometric redshifts, at least in the LSB/dwarf regime, compared to what is possible using previous datasets. 

\begin{figure*}
\center
\includegraphics[width=0.9\textwidth]{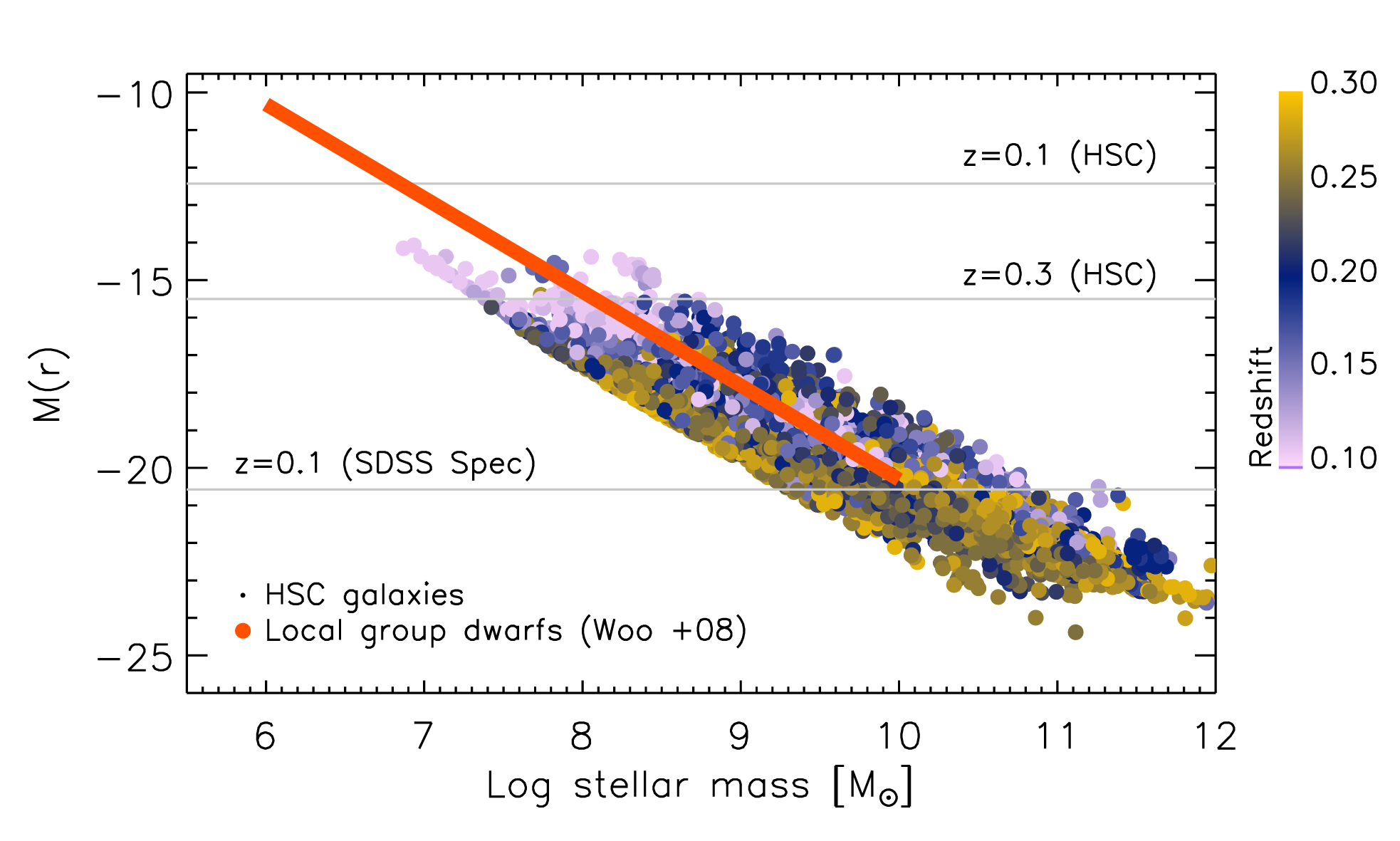}\\
\includegraphics[width=0.9\textwidth]{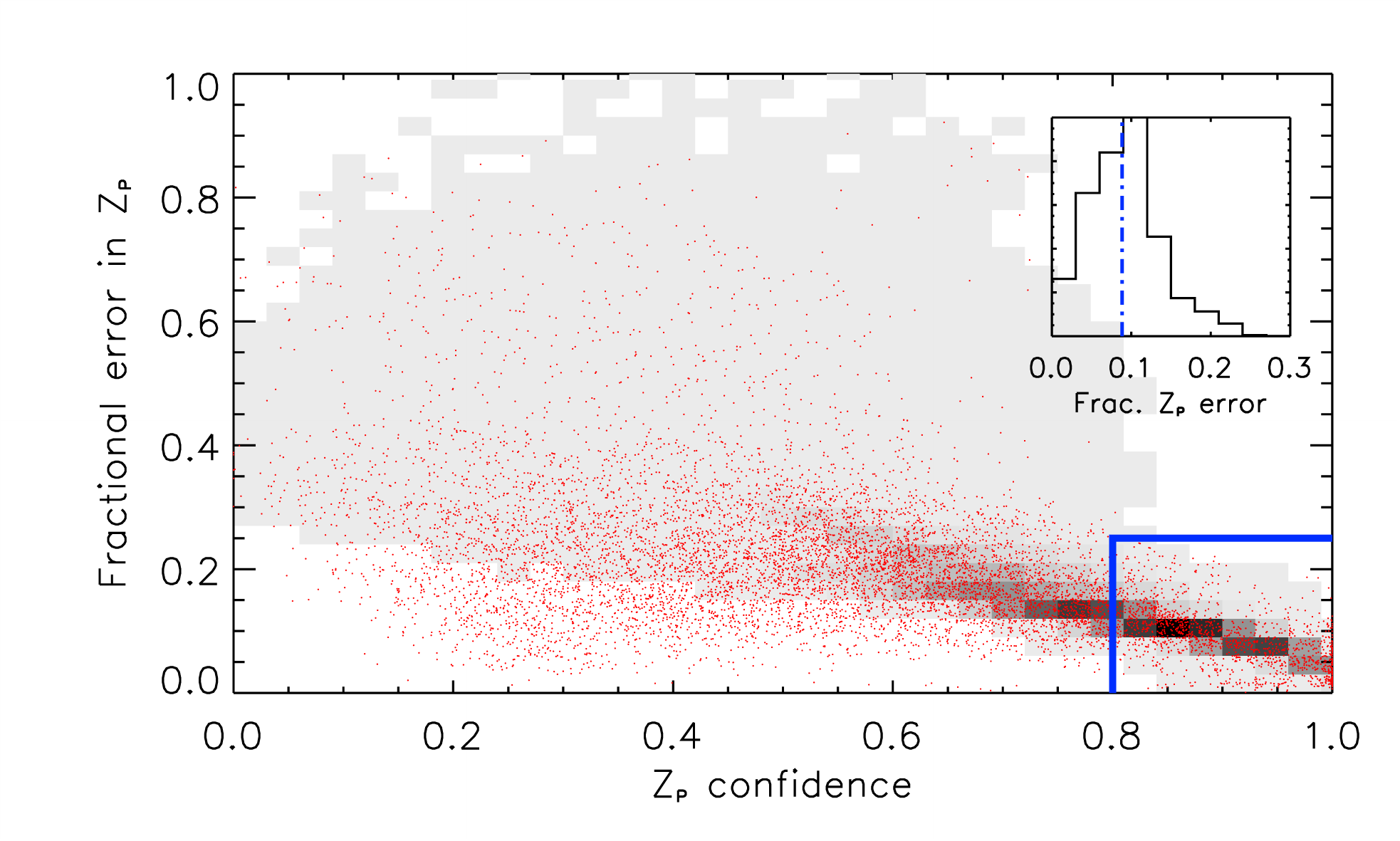}
\caption{\textbf{Top:} Completeness limits in the Wide layer HSC-SSP DR1, which is $\sim$4 mags deeper than the SDSS. This is similar in depth to data from the LSST commissioning surveys, with the full LSST survey being 2 mags deeper still over half the sky. By virtue of its deeper imaging, the HSC-SSP DR1 is able to detect \emph{standard} dwarfs (M* $<$ 10$^9$ M$_{\odot}$) at \textit{cosmological} distances. \textbf{Bottom:} Accuracy of photometric redshifts in the Wide layer of the HSC-SSP DR1. Black dots indicate massive galaxies, while red dots indicate dwarfs. When the redshift confidence parameter (Z$_p$) is greater than 0.8, the fractional photometric redshift errors are better than 10 per cent in the dwarf regime. Reproduced from Figure 1 in \citet{Kaviraj2019}.}
\label{fig:science with photozs}
\end{figure*}

The top panel of Figure \ref{fig:science with photozs} (taken from \citet{Kaviraj2019}) indicates completeness limits, at various redshifts, in the Wide layer of the HSC-SSP DR1, which is $\sim$4 mags deeper than the SDSS. This is similar in depth to data expected from forthcoming LSST commissioning surveys, with the full LSST survey (expected in the early 2030s) being a further 2 mags deeper over around half the sky. It is worth noting that, by virtue of its deeper imaging, the HSC-SSP DR1 is able to detect \emph{standard} dwarfs (M* $<$ 10$^9$ M$_{\odot}$) at \textit{cosmological} distances. This is difficult to do, for example, in the SDSS spectroscopic sample which is significantly shallower, so that dwarfs which appear in this dataset outside the local Universe are likely to be anomalously bright (e.g. due to unusually high levels of star formation). 

The bottom panel of Figure \ref{fig:science with photozs} shows the accuracy of photometric redshifts in the Wide layer of the HSC-SSP DR1, calculated using the \texttt{MIZUKI} code \citep{Tanaka2015}. Black dots indicate massive galaxies, while red dots indicate dwarfs. When the redshift confidence parameter (Z$_p$) is greater than 0.8, the fractional photometric redshift errors are better than 10 per cent in the dwarf regime. It is worth noting here that similar photometric-redshift accuracies (i.e. around 10 per cent) have been routinely used to study massive galaxies at intermediate and high redshift. It will, therefore, be possible to perform robust science, at least in a statistical way, in the LSB/dwarf regime from deep-wide surveys using photometric redshifts alone. 
While data from the HSC-SSP DR1 indicates that photometric redshifts can indeed be used to perform statistical studies using deep-wide surveys, continued investment in improving photometric-redshift pipelines, using spectroscopic surveys like WAVES as benchmarks, is desirable, in order to maximize the scientific potential of these datasets.

\section{Summary}

The advent of deep-wide optical surveys, from instruments like the HSC, LSST and EUCLID, are poised to open up a revolutionary new frontier in extra-galactic astrophysics. These surveys will offer the first truly statistical view of galaxies in the LSB/dwarf regime -- which dominate the galaxy number density -- in all environments and across cosmic time. They will also enable the first statistical studies of key LSB structures, that are largely invisible in past wide-area datasets, but offer strong constraints on our structure-formation paradigm, such as merger-induced LSB tidal features and ICL. 

However, while revolutionary advances are expected from these surveys, some data-analysis challenges must be addressed to enable us to extract the full potential of these datasets. Data-processing pipelines must preserve LSB structures, which are notoriously susceptible to sky over-subtraction and shredding by de-blenders. The prodigious volumes of data expected will require machine-learning methods (in particular, \textit{unsupervised} techniques) to augment or even replace traditional methods of analysis. New cosmological simulations, such as New Horizon and Illustris-TNG50, that have high mass and spatial resolution, will be important for both making statistical predictions in the LSB/dwarf regime and providing spatially-resolved mock data for more massive galaxies. Nevertheless, since the increase in spatial resolution comes at the cost of being able to simulate large volumes, lower-resolution simulations with larger box sizes and semi-analytical models will remain valuable theoretical tools.

Finally, while these new surveys will enable us to explore fainter and lower-mass galaxies, the derivation of physical properties (e.g. stellar masses, rest-frame colours and star formation rates) will require reliable redshift information. Since even the next generation of spectrographs are unlikely to provide spectroscopic redshifts in the LSB/dwarf regime beyond the local Universe, photometric redshifts are likely to underpin a significant fraction of science using these facilities. While the HSC-SSP DR1 indicates that photometric redshifts in the LSB/dwarf regime can indeed facilitate statistical studies, continued investment in photometric redshift pipelines, benchmarked against new spectroscopic surveys, is needed to enable us to fulfil the scientific potential of these new datasets. 

As these proceedings indicate, these data-analysis challenges are tractable - the convergence of deep-wide surveys, machine-learning techniques and high-resolution cosmological simulations is, therefore, poised to make a transformational impact on our understanding of galaxy evolution in the coming years. 


\section*{Acknowledgements}
I thank Yohan Dubois, Lee Kelvin, Alex Hocking, Garreth Martin and Sukyoung Yi for providing figures included in these proceedings prior to their publication. I acknowledge generous financial support from the LOC/SOC and from the STFC for UK participation in LSST through grant ST/N002512/1. 



\end{document}